# Q-MRS: A Deep Learning Framework for Quantitative Magnetic Resonance Spectra Analysis

Christopher J. Wu[1], Lawrence S. Kegeles[2], Jia Guo[2]
[1] Taub Institute for Research on Alzheimer's Disease and the Aging Brain,
Columbia University, New York, NY, USA
[2] Department of Psychiatry, Columbia University, New York, NY, USA

*Abstract*—**Magnetic resonance spectroscopy (MRS) is an established technique for studying tissue metabolism, particularly in central nervous system disorders. While powerful and versatile, MRS is often limited by challenges associated with data quality, processing, and quantification. Existing MRS quantification methods face difficulties in balancing model complexity and reproducibility during spectral modeling, often falling into the trap of either oversimplification or over-parameterization. To address these limitations, this study introduces a deep learning (DL) framework that employs transfer learning, in which the model is pre-trained on simulated datasets before it undergoes fine-tuning on in vivo data. The proposed framework showed promising performance when applied to the Philips dataset from the BIG GABA repository and represents an exciting advancement in MRS data analysis.**

## INTRODUCTION

Magnetic resonance spectroscopy (MRS) is a noninvasive technique that has been widely used for the metabolic profiling of the central nervous system in a localized manner. Despite its utility, challenges relating to data acquisition and processing often limit the amount of useful information that can be obtained from MRS data. One main issue is the relatively low sensitivity of MRS, as it detects metabolites in the millimolar range against a backdrop of molar-range tissue or free water signals. This results in a low signal-to-noise ratio (SNR) in the data that can be increased to a certain extent by acquiring more transients at the cost of longer scan times. In addition, variability in vendor-provided software contributes to non-standardized acquisition outcomes, which further necessitates systematic data processing and analysis to enable accurate interpretation of the MRS data [1], [2]. Fortunately, many third-party software are available for comprehensive spectral processing and analysis. Examples of these programs include open-source tools like Tarquin [3] and Vespa [4], academic software such as jMRUI [5] and INSPECTOR [6], as well as commercial packages like LCModel [7].

After pre-processing, a major step in MRS data analysis is the fitting of the spectra for metabolite quantification. In general, LCM is the favored method for spectral fitting thanks to its effectiveness and adaptability, [2]. However, traditional LCM approaches are sensitive to noise and other artifacts in the spectral data. Moreover, methods that have a large number of parameters in the parametric equation can easily overfit the spectra if not adequately regularized. To address the issue of over-parameterization, traditional programs have introduced soft and hard constraints [8]. Soft constraints that are imposed on metabolite amplitude ratios inevitably embed assumptions into the model and may not always accurately reflect the true underlying metabolite distributions. A poorly chosen constraint can inflate reproducibility while leading to inaccurate quantification results. On the other hand, deep learning (DL) approaches for metabolite quantification often neglected critical parameters, such as metabolite-specific line-broadening and first-order phase shift [9], [10], [11], [12], [13].

The oversimplification in previous DL methods can be addressed by incorporating all the necessary parameters. In addition, layer freezing during transfer learning can help constrain the DL model to deal with the problem of over-parameterization. With these considerations, the authors aim to overcome the limitations of current quantification methods by introducing Q-MRS (Fig. 1), a DL framework that involves pre-training the model on simulated data before fine-tuning it on the in vivo data.

## MATERIALS AND METHODS

*Basis Set*

A basis set containing fourteen metabolites of interest was compiled in Osprey [14] and simulated using the FID-A software [15]. with the same scanning parameters as those of in vivo data. The metabolites included are aspartate (Asp), creatine (Cr), phosphocreatine (PCr), gamma-aminobutyric acid (GABA), glutamate (Glu), glutamine (Gln), N-acetylaspartate (NAA), N-acetylaspartylglutamate (NAAG), glutathione (GSH), phosphocholine (PCh), glycerophosphorylcholine (GPC), myo-inositol (mI), scyllo-inositol (sI), and taurine (Tau). Additionally, three macro-molecule (MM) basis functions were included: MM20, Lip20, and MM3co, a co-edited MM signal at 3.0 ppm for the difference (DIFF) spectra. Lastly, -CrCH2 serves as a correction term to account for the effects of water suppression and relaxation on the Cr CH2 singlet. Therefore,

Corresponding author: J. Guo (email: jg3400@columbia.edu).

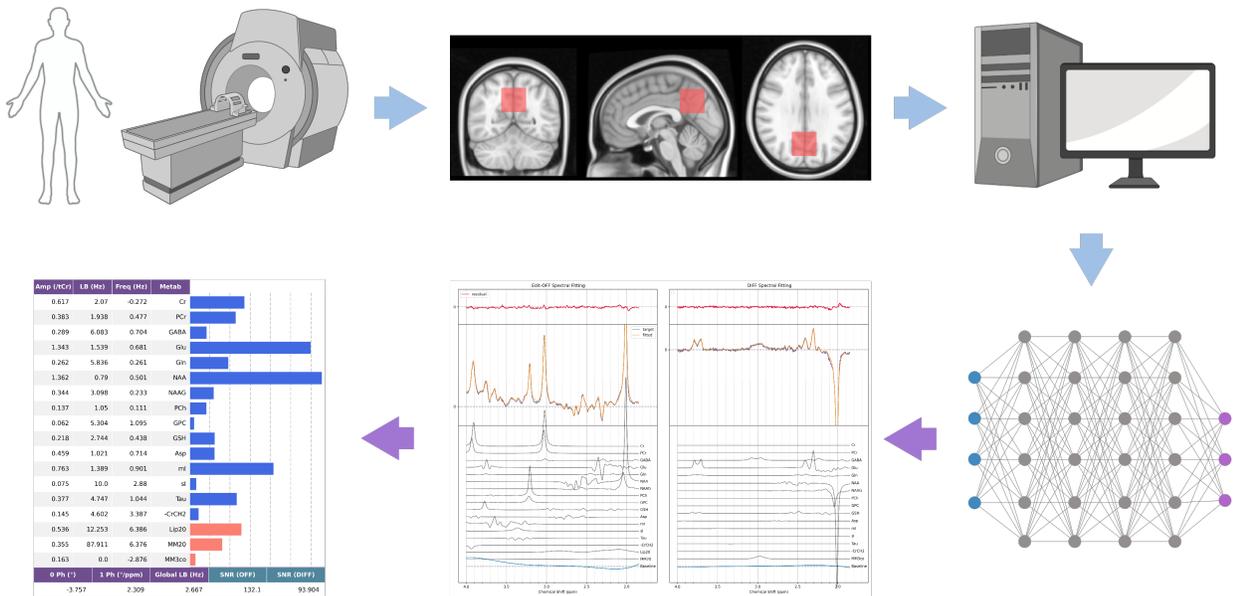

Fig. 1: Overall pipeline of MRS data from acquisition to metabolite quantification by Q-MRS. The pipeline begins with data acquisition, where scans are performed to collect raw spectral data from the human brain. Following the acquisition, the data undergoes pre-processing steps to enhance signal quality. The pre-processed spectra are then subjected to LCM via Q-MRS to estimate the concentrations of various metabolites.

the basis set contains a total of eighteen basis functions. A figure of the simulated basis set (Fig. S1) is shown in the Supplementary Material.

### *In Vivo Dataset*

GABA-edited MEGA-PRESS [16], [17] datasets from the Big GABA repository [18], acquired on a 3T Philips scanner, were collected from subjects (age range: 18-34 years, 52 males, 49 females). The dataset contains 101 pairs of edit-ON/-OFF spectra with the following scan parameters: TR/TE = 2000/68 ms, 320 averages, and editing at 1.9/7.46 ppm for edit-ON/-OFF, respectively.

The coil-combined data were pre-processed using Osprey's *Proc* (Process) module, which involved multiple operations including spectral registration, eddy-current correction, and linear baseline correction.

### *Simulated Dataset*

The simulated dataset contains 100,000 pairs of edit-OFF and DIFF spectra. The amplitude ranges for each metabolite were obtained from a meta-analysis of healthy young adults aged 25 to 35 [19].. Additional parameters such as the global line-broadening factor and phases were derived from an initial fitting of the in vivo dataset, as described in the upcoming section. The minimum and maximum values of the metabolite amplitudes are summarized in Table S1 in the Supplementary Material.

The spectra were simulated from a uniform distribution with added randomness using Sobol sequences for each parameter. The parameters were sampled within their respective minimum and maximum values to ensure comprehensive coverage of the parameter space. The simulated spectra were constructed using LCM in the same manner as described in the following section on model architecture, with the only difference being the addition of Gaussian noise prior to normalization of the spectra.

### *Framework*

The Q-MRS framework can be summarized in three major steps:

1) **Initial Fitting of In Vivo Data**
   During the initial fitting stage, an untrained model was used to fit the in vivo data to obtain approximate ranges of values for parameters besides the metabolites amplitudes. Two soft constraints on metabolite amplitude ratios were added to the loss function (Table S3 in the Supplementary Material) to stabilize the model and ensure more reasonable predictions. These constraints were removed during the final fitting of in vivo data.

2) **Pre-Training on Simulated Data**
   The model was trained on the simulated dataset with a train/validation/test split of 8:1:1 using the mean squared error (MSE) between the actual and predicted values of the parameters as the loss function. Two separate models were evaluated on the simulated dataset: a CNN model and a CNN-LSTM model. The two models are identical except for the inclusion of a LSTM layer in the latter. Training was performed using the Adam optimizer with a batch size of 256 and an initial learning rate of 0.005. Training details on early stopping and learning rate schedule are described in the Supplementary Material.

3) **Fine-Tuning and Final Fitting of In Vivo Data**
   The pre-trained model had all but its final output layers in each head of the multi-layer perceptron (MLP) frozen and was then fine-tuned on in vivo data to generate the final predictions. During transfer learning, no constraints on metabolite amplitude ratios were imposed. Training was performed using the Adam optimizer with a batch size of 1 and an initial learning rate of 0.0001. The training was stopped early when no improvement was observed in 40 epochs.

*Network Architecture*

The network architecture is detailed in Fig. 1 and can be summarized as an CNN-LSTM model with a multi-headed MLP. Each head of the MLP is responsible for predicting a different set of parameters: metabolite amplitudes, global parameters (Gaussian line-broadening, zeroth- and first-order phase), individual (metabolite-specific) Lorentzian line-broadening, individual frequency shifts, and polynomial baseline coefficients. A list of hard constraints imposed on the parameters are presented in Table S2 in the Supplementary Material.

$$s(t) = \left( \sum_{m=1}^{M} A_m B_m(t) \cdot e^{2\pi i (F_m t)} e^{-\pi L_m t} \right) e^{-(\pi G t)^2} e^{2\pi i \left( \frac{\phi_0}{360} \right)} \quad (1)$$

$$S(f) = FFT(s(t)) e^{2\pi i (\phi_1 f)} + b(f) \quad (2)$$

From the predicted parameters, the network constructs the predicted spectra using Eq. 1 and Eq. 2, where $s(t)$ and $S(f)$ are the time-domain and frequency-domain signals, respectively. $A_m$, $B_m$, $F_m$, and $L_m$ are the scaling factor (amplitude), basis function, the frequency shift, and Lorentzian damping (line-broadening) factor for the m-th metabolite or MM free-induction decay (FID). $G$, $\phi_0$, and $\phi_1$ refer to the Gaussian line-broadening factor, zeroth-order phase, and first-order phase, respectively. After fast Fourier transform (FFT), two separate fifth-degree polynomials $b(f)$ baselines are added for the edit-OFF and DIFF spectra.

*Measurement of Fitting Quality*

Three goodness-of-fit metrics were used to evaluate the fitting quality on the in vivo dataset: the MSE between the fitted (predicted) and target spectra, the standard deviation (SD) of the fit residual, and the fit quality number (FQN), which is defined as the ratio of the variance in the fit residual to the variance in the pure spectral noise [21]. Lower values of MSE, SD of the fit residual, and FQN all indicate better fitting quality. For an ideal fit, the FQN should be close to 1 [2].

|      | DeepMRS-net | Q-MRS CNN | CNN-LSTM |
|------|-------------|-----------|----------|
| Cr   | 1.51        | 1.39      | **1.38** |
| Pcr  | 3           | 2.55      | **2.46** |
| GABA | 2.1         | **2.05**  | **2.05** |
| Glu  | 1.33        | 1.3       | **1.28** |
| Gln  | 4.1         | 4.03      | **3.93** |
| NAA  | 1           | 0.97      | **0.95** |
| NAAG | 5.7         | 5.36      | **5.14** |
| PCh  | 3.16        | 3.13      | **3.09** |
| GPC  | 2.43        | **2.37**  | **2.37** |
| GSH  | 3.69        | 2.64      | **2.62** |
| Asp  | 5.37        | 5.37      | **5.35** |
| mI   | 1.5         | 1.49      | **1.48** |
| sI   | 12.44       | 10.73     | **10.38**|
| Tau  | 7.42        | 7.46      | **7.35** |

TABLE I: Performance comparisons on simulated test dataset. For each metabolite, the average MAPE across all the spectra in the dataset is reported.

*Reproducibility and Robustness Evaluation*

To evaluate reproducibility, the in vivo dataset was fitted in two separate runs. Additionally, to access the model's robustness, Gaussian noise was added to the spectra at two different levels, reducing the SNR by factors of $\sqrt{2}$ and 2, thereby creating two noisy datasets. This addition of noise simulates scenarios where the number of acquired transients is halved and quartered, respectively. The SNR was calculated by dividing the amplitude of the NAA peak at around 2 ppm by the SD of the detrended noise in the spectrum from -2 to 0 ppm.

## RESULTS

*Simulated Dataset*

The mean absolute percentage error (MAPE) between the predicted metabolite amplitudes and the ground truth for three network architectures - DeepMRS-net, a CNN, and a CNN-LSTM proposed in this work - using the Q-MRS framework is summarized in Table 1. The DeepMRS-net evaluated here is a slightly modified version of the original model introduced in the abstract [11], and the model architecture used in this paper is detailed in Fig. 2 in the Supplementary Material. The Q-MRS framework utilizing the CNN-LSTM network demonstrated the lowest MAPE across all metabolites. It is tied with Q-MRS CNN for GABA and GPC. The following results will focus on the CNN-LSTM version.

*In Vivo Dataset Fitting Quality*

Table 2 summarizes the three metrics used to assess spectral fitting quality, measured using data points from 1.85 to 4.0 ppm in the spectra. Among the three software, LCModel shows the lowest scores for all metrics for the fitting of the edit-OFF spectra, while Q-MRS shows the lowest scores for the fitting of the DIFF spectra.

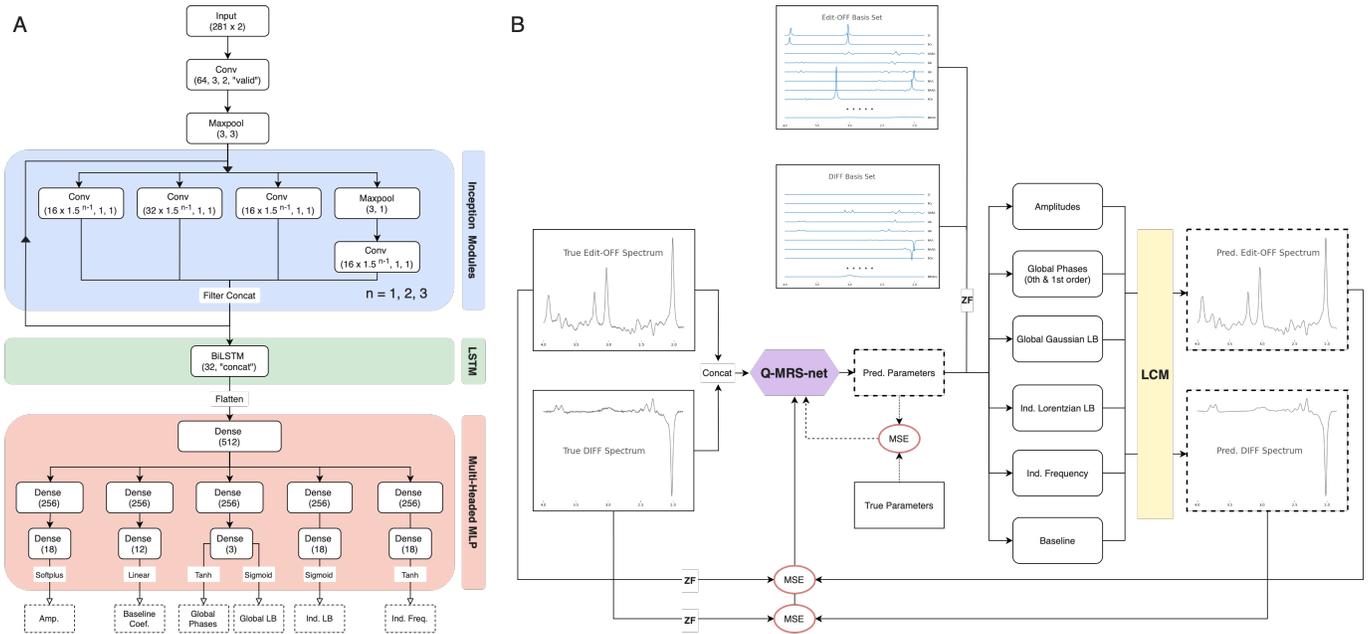

Fig. 2: Q-MRS framework and CNN-LSTM architecture. **(A)** The network architecture consists of a 1D convolution layer and max pooling layer followed by three inception modules [20] stacked sequentially, where *n* determines the number of filters and increases by 1 after each module, followed by a bidirectional LSTM using concatenated merge mode. A multi-headed MLP completes the architecture. A rectified linear unit (ReLU) activation function is applied after each convolution and hidden fully connected layer. **(B)** The input data to the Q-MRS-net is a 281 x 2 tensor, where the row represents the data points in the real component of a spectrum and the columns represent the channel size corresponding to the edit-OFF and DIFF spectra. The model outputs predicted values corresponding to metabolite amplitudes, line-broadening factors, phase shifts, frequency shifts, and baseline. Predicted spectra are constructed from these values through LCM using zero-filled basis functions. After the imaginary component is discarded, the predicted spectra are cropped to be in the same ppm range as the input and normalized. The loss function is MSE between the output predicted spectra and zero-filled input target (true) spectra. Alternatively, for training with simulation data, the loss is the MSE between the output predicted parameters and the ground truth.

|  |  | LCModel | Osprey | Q-MRS |
|---|---|---|---|---|
| edit-OFF | MSE (x10,000) | **0.81** | 2.76 | 2.3 |
|  | SD of residual | **0.0087** | 0.016 | 0.014 |
|  | FQN | **5.73** | 18.96 | 12.58 |
| DIFF | MSE (x10,000) | 6.1 | 10.95 | **1.65** |
|  | SD of residual | 0.013 | 0.017 | **0.012** |
|  | FQN | 19.84 | 28.25 | **5.26** |

TABLE II: Fitting quality comparisons on the in vivo data set. For each metric, the average value across all the spectra in the dataset is reported. The metrics for the fitting of edit-OFF spectra and the fitting of DIFF spectra are reported separately.

*Estimated Metabolite Levels*

For Q-MRS, the mean values of tCr ratios of tNAA, tCho, mI, and Glx for all subjects in the dataset were 1.46 ± 0.07, 0.18 ± 0.01, 0.75 ± 0.07, and 1.38 ± 0.07, respectively.

The estimated metabolite levels from the three models are presented in Fig. 3. The mean estimated values for GABA+/tCr are 0.39 ± 0.12, 0.35 ± 0.11, and 0.39 ± 0.075 for LCModel, Osprey, and Q-MRS, respectively. GABA/tCr Q-MRS shows the highest Glu and Gln levels, and as as a results, a highest Glx level between the methods. The Glu/Gln for the three methods are 10.50, 5.71, and 3.97. The tCho levels between the software show good agreement but are significantly lower than the values reported in the meta-analysis. A box plot showing the same estimated metabolite levels (Fig. S3) is shown in the Supplementary Material.

*Reproducibility and Robustness of Q-MRS*

Table 3 shows PCC between each pair of runs for estimated GABA+, Glx, tNAA, and tCho amplitudes. An example fit of a spectrum with SNR reduced by half (Fig. S4), including contributions from individual metabolites, is presented in the Supplementary Materials.

*Q-MRS Subject Quantification Report*

Fig. 4 shows the Q-MRS quantification report for one subject in the original in vivo dataset without any added noise.

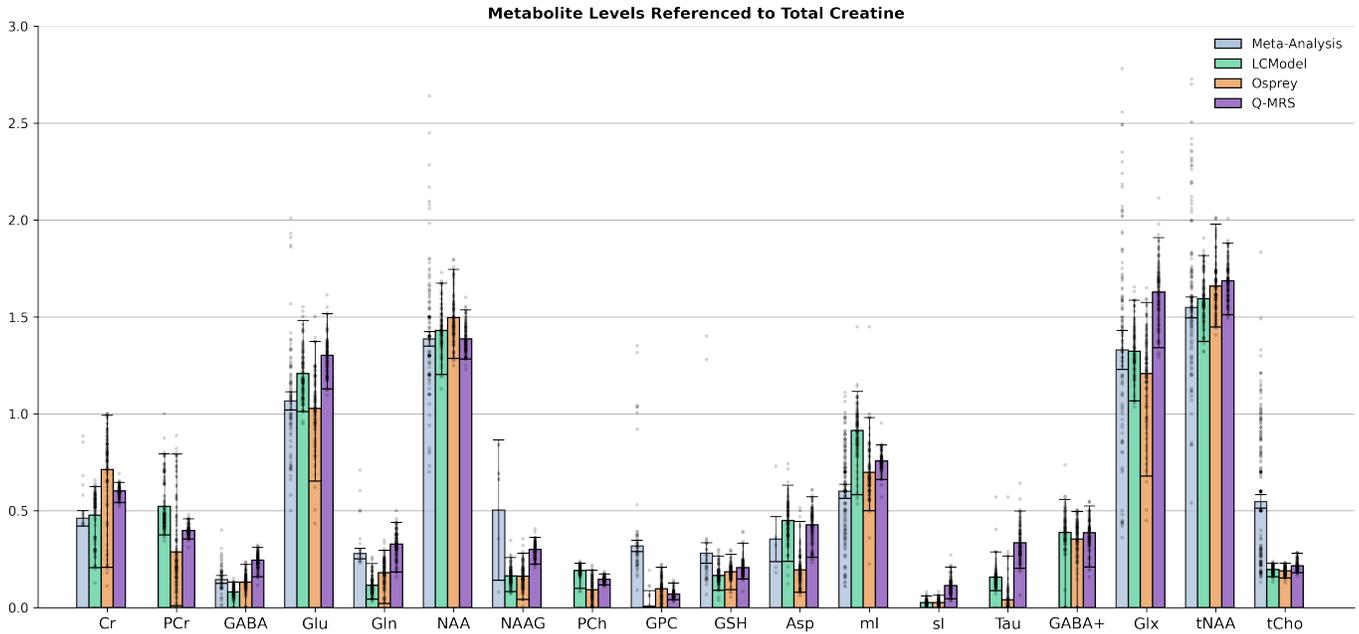

Fig. 3: Estimated metabolite amplitudes referenced to total creatine. The mean estimated metabolite amplitudes referenced to total creatine (tCr) for the three software packages across all 101 subjects are plotted and compared with those reported in a meta-analysis of healthy younger adults (18–45 years). The lines on the bars represent the interquartile range (IQR), with outliers indicated as dots.

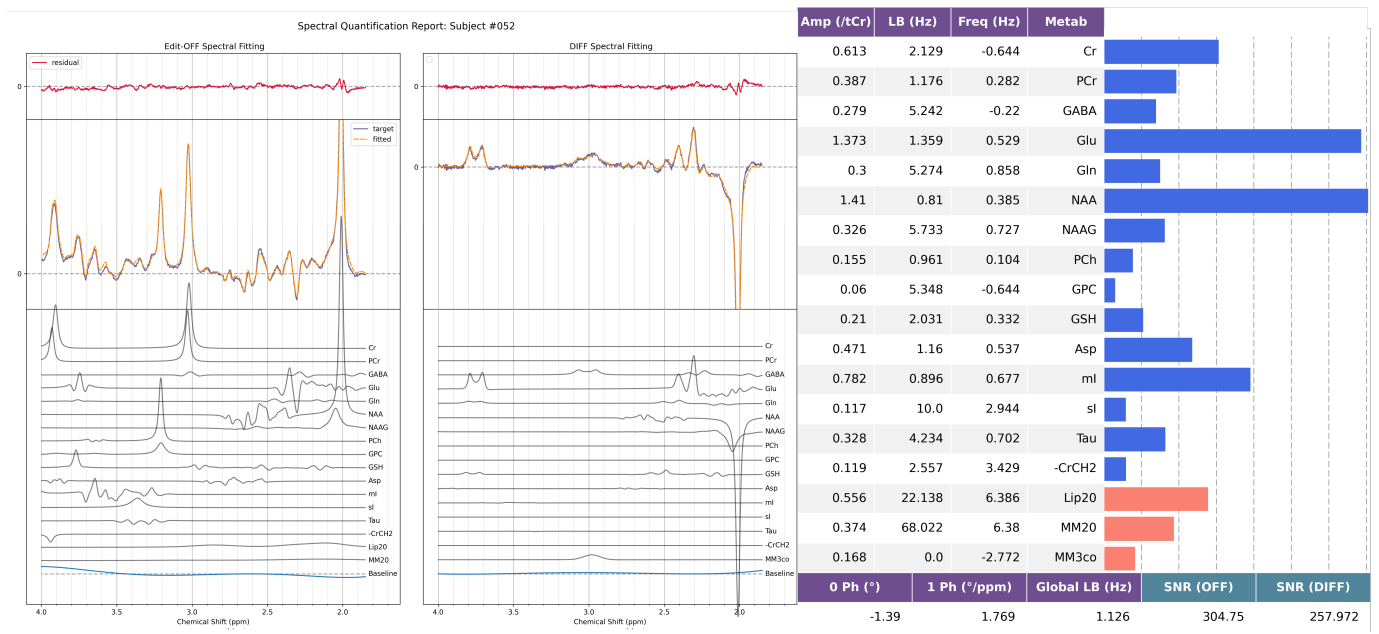

Fig. 4: Example of Q-MRS quantification report for a subject. The spectral fitting plot is displayed on the left, with the predicted parameters on the right. On the right of the table with metabolite-specific parameters is a bar graph displaying the metabolite levels. At the bottom, the predicted global parameters and SNRs for the subspectra are presented.

| Run Pair | PCC (GABA+, Glx, tNAA, tCho) |
| --- | --- |
| Orig 1 vs Orig 2 | 1.00, 1.00, 1.00, 1.00 |
| Orig 1 vs Noisy 1 | 0.99, 0.99, 0.99, 0.99 |
| Orig 1 vs Noisy 2 | 0.98, 0.96, 0.99, 0.98 |
| Orig 2 vs Noisy 1 | 0.99, 0.99, 0.99, 0.99 |
| Orig 2 vs Noisy 2 | 0.98, 0.96, 0.99, 0.98 |
| Noisy 1 vs Noisy 2 | 0.97, 0.97, 0.99, 0.98 |

TABLE III: PCC of metabolite amplitudes between different runs. "Orig 1" and "Orig 2" refer to runs on the original in vivo dataset without additional noise. "Noisy 1" and "Noisy 2" corresponds to the runs with added noise, reducing SNR by factor of $\sqrt{2}$ and 2, respectively.

## DISCUSSION

The comparative analysis between LCModel, Osprey, and Q-MRS shows that the mean estimated values for key metabolite concentrations such as GABA+, tNAA, and tCho are generally consistent across these tools. However, while the GABA+ levels align well across the software, the individual GABA levels differ greatly. For Osprey and Q-MRS, the mean GABA/GABA+ values are 0.38 and 0.64, respectively. This disparity points to The inherent difficulty in separating the GABA and co-edited MM signals at 3T, a challenge well-documented in the literature [{]MM3co, [{]MM3co2. Previous studies have reported that the MM resonance accounts for approximately 50% or more of the total signal at around 3.0 ppm [{]50-1993, [26], [27]. The relative contributions of MM and GABA may depend on the voxel selection [24]. Additionally, during spectral fitting, these values could vary due to the constraints applied to the ratio of MM3co to GABA and the stiffness of the baseline, as a flexible baseline might partially account for MM, resulting in a higher relative contribution of GABA compared to the co-edited MM [27].

The distinction between the methods themselves warrants further discussion. LCModel and Osprey utilize non-machine-learning algorithms while Q-MRS employs a DL approach. Unlike traditional optimization methods that focus on minimizing the residuals between the target and predicted spectra, DL models make predictions based on features extracted from the data. The ability to learn and generalize from data makes them well-suited for MRS quantification, even in low SNR scenarios.

The model proposed in this study uses convolutional layers to extract key spectral features, which are then processed by an LSTM layer. The LSTM layer may help the model distinguish between overlapping peaks, such as those of Glu and Gln at around 3 ppm, which are difficult to separate \cite{glugln}. However, by integrating information across the spectrum, the metabolites can be more easily separated due to their distinct upfield resonances. The LSTM can remember patterns that span across different parts of the spectrum, enabling the model to contextualize local features within a broader perspective. This multi-layered approach takes advantage of both local and global spectral features, resulting in more consistent and reliable quantification as indicated by a reduced variation in predicted metabolite levels across the subjects.

To address the problem of over-parameterization, the Q-MRS framework incorporates several strategic constraints. The first set of constraints pertains to J-difference editing sequences like MEGA-PRESS. In Q-MRS, the simultaneous fitting of the edit-OFF and DIFF spectra involves sharing parameters between the two spectral modes, which prevents the model from overfitting to any single spectrum. This is similar to the "concatenated fitting" option available in Osprey, where the edit-OFF and SUM spectra are fitted together. The second set of constraints is imposed on the values of phases, line-broadening factors, and frequency shifts. These constraints are implemented through squashing functions such as the sigmoid and hyperbolic tangent functions, forcing the parameters to remain within physically plausible ranges. The third constraint is introduced during the fine-tuning phase, when all layers up to the final output layer in each MLP head are frozen, leaving only a small subset of model parameters to be trainable. This significantly constrains the model while still allowing it to adapt to the specific characteristics of the new data.

Zero-filling the time-domain signal before FFT serves an important purpose when performing frequency shifts of the spectrum. The precision of a frequency shift is limited by the frequency resolution, which is defined as the sampling rate divided by the number of points in the FFT. In the time domain, multiplying the signal with an exponential decay results in a frequency shift in the spectrum. Here, the damping factor must be an integer multiple of the frequency resolution. This presents a challenge for backpropagation as gradients cannot be computed for functions involving integer-type tensors. To address this, the FIDs are zero-filled to reduce the frequency granularity of the spectra. This minimizes the unintended effects of fractional frequency shifts on the spectral appearance.

While Q-MRS showed promising results, a limitation of this study is the lack of diversity in the in vivo datasets being evaluated. In the future, the framework should be tested on more datasets varying in vendors, sequences, brain regions, and other factors.

## CONCLUSION

In conclusion, this study presents a DL framework that effectively addresses some of the problems faced by existing methods of MRS data analysis. For future work, more advanced and sophisticated model architectures should be explored. Notably, transformers and diffusion models have had a tremendous impact across many machine learning domains. These models could also find success in the field of MRS. The authors believe that through continued investigation and development, future DL methods could achieve unprecedented levels of accuracy and reliability, opening new avenues for clinical and research applications.

## ACKNOWLEDGEMENT

This research was generously supported by grants from the Simons Foundation and The University of Texas MD Anderson Cancer Center. The authors extend their heartfelt gratitude to Dr. Georg Oeltzschner and Dr. Richard Edden for their invaluable guidance and support in utilizing the Osprey software.

# Supporting Information for

**Q-MRS: A Deep Learning Framework for Quantitative Magnetic Resonance Spectra Processing and Analysis**

Christopher J. Wu, Lawrence S. Kegeles, Jia Guo

Jia Guo.
E-mail: jg3400@columbia.edu

**This PDF file includes:**

    Figs. S1 to S4
    Tables S1 to S3

**Simulated Dataset Training Details.** During training on the simulated dataset, the validation MSE loss was monitored, and the learning rate was reduced to 80% of the original when no improvement was observed in 30 epochs. When the learning rate was reduced to 20% of its initial value, it was not decreased further. The training was stopped early when no improvement was observed in 40 epochs.

**Christopher J. Wu, Lawrence S. Kegeles, Jia Guo**

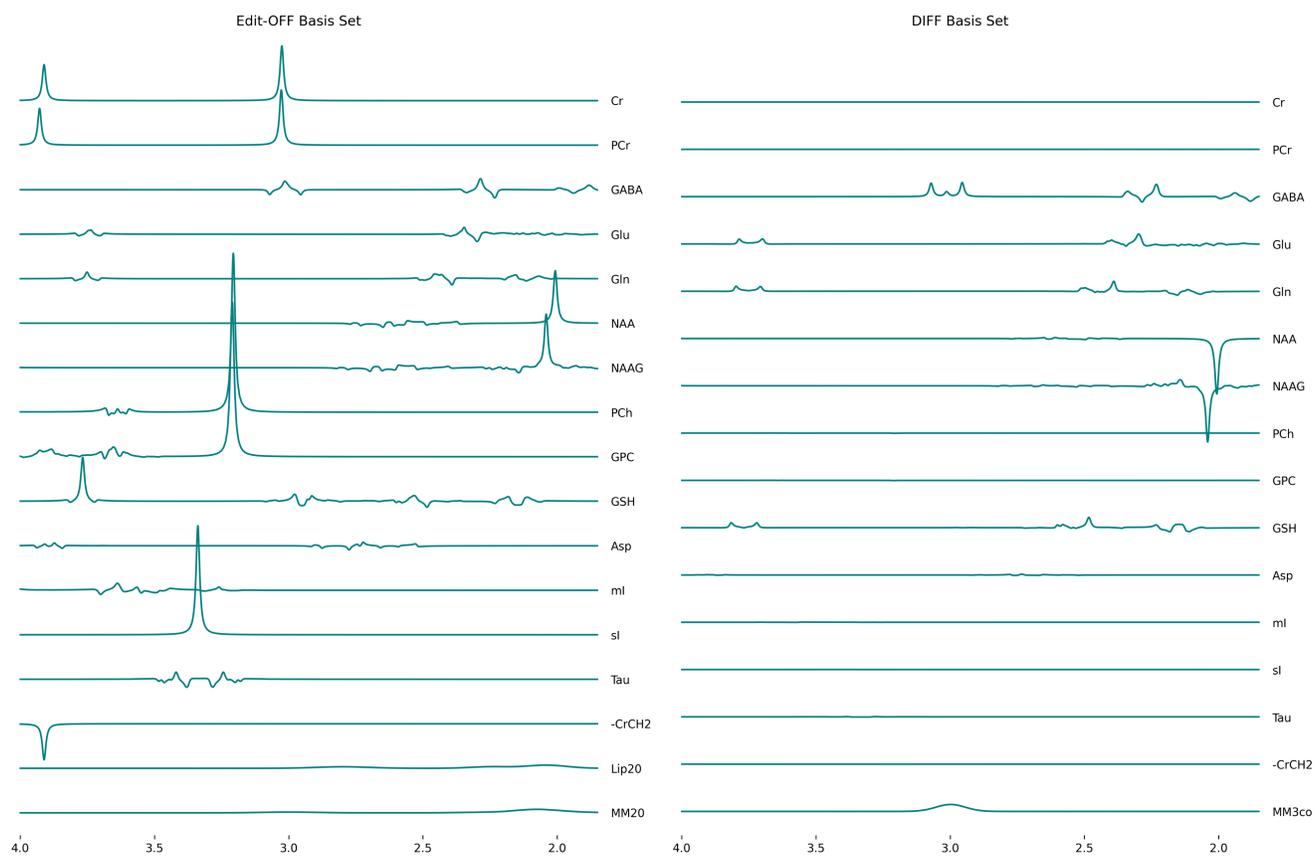

**Fig. S1.** Basis set used for simulation dataset and LCM fitting.

|      | Min   | Max   |
|------|-------|-------|
| Cr    | 5.49  | 6.42  |
| Pcr   | 3.43  | 5.01  |
| GABA  | 2.03  | 2.22  |
| Glu   | 8.45  | 9.14  |
| Gln   | 2     | 2.41  |
| NAA   | 10.43 | 11.03 |
| NAAG  | 1.42  | 1.93  |
| PCh   | 0.58  | 1.17  |
| GPC   | 1.55  | 1.87  |
| GSH   | 1.3   | 1.57  |
| Asp   | 1.94  | 2.69  |
| mI    | 5.58  | 6.08  |
| sI    | 0.14  | 0.22  |
| Tau   | 1.02  | 1.49  |
| MM3co | 0     | 4.43  |

**Table S1.** The minimum and maximum values used for the metabolite amplitudes to simulate data. The values are sampled using a pseudo-random uniform distribution defined by the ranges of these values. The maximum value of MM3co is twice that of GABA.

| Parameter | Min | Max |
|---|---|---|
| $\phi_0$ | -360° | 360° |
| $\phi_1$ | -10°/ppm | 10°/ppm |
| $G$ | 0 Hz | $\sqrt{5000}$ Hz |
| $L_{metab}$ | 0 Hz | 10 Hz |
| $L_{MM/Lip}$ | 0 Hz | 100 Hz |
| $f_{metab}$ | -0.03 ppm | 0.03 ppm |
| $f_{MM/Lip}$ | -0.05 ppm | 0.05 ppm |

**Table S2.** Hard constraints on the parameters are imposed to stabilize the solution and are defined as they are in LCModel, Tarquin, and Osprey (Table S3). Here, $\phi_0$, $\phi_1$, $G$, $L$, and $f$ represent the global zero-order phase correction; the global first-order (linear) phase correction, the global Gaussian line-broadening factor, the individual Lorentzian line-broadening factors and the individual frequency shifts, respectively.

| Ratio | Target Value | Tolerance |
|---|---|---|
| NAAG/NAA | 0.15 | 0.15 |
| GABA/BIG3 | 0.04 | 0.04 |

**Table S3.** The following soft constraints on metabolite ratios are only used during the initial fitting stage. BIG3 is defined as tNAA + tCr + tCho + tCho + tCho as it is in LCModel. If the difference between the predicted value of the metabolite ratio and its target value by more than the tolerance, a penalty is added to the loss.

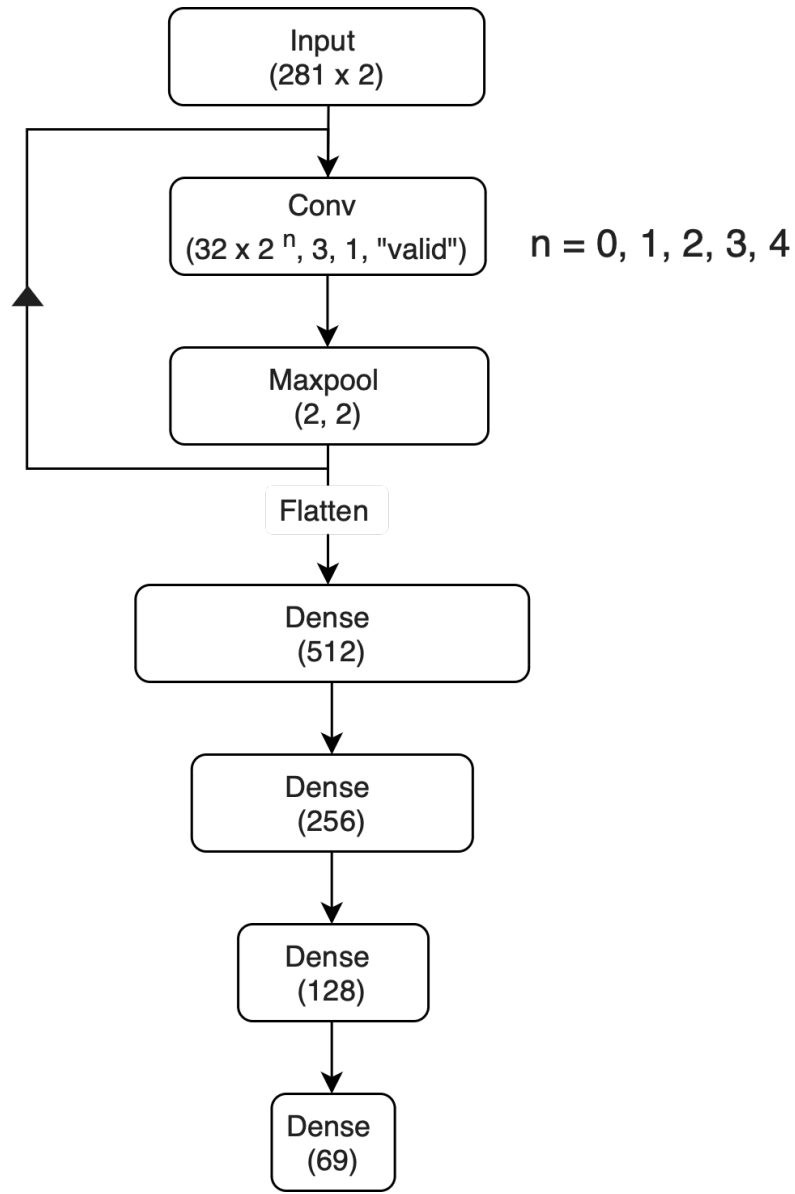

**Fig. S2.** DeepMRS-net architecture. A ReLU activation function is applied after each convolution and hidden fully connected layer.

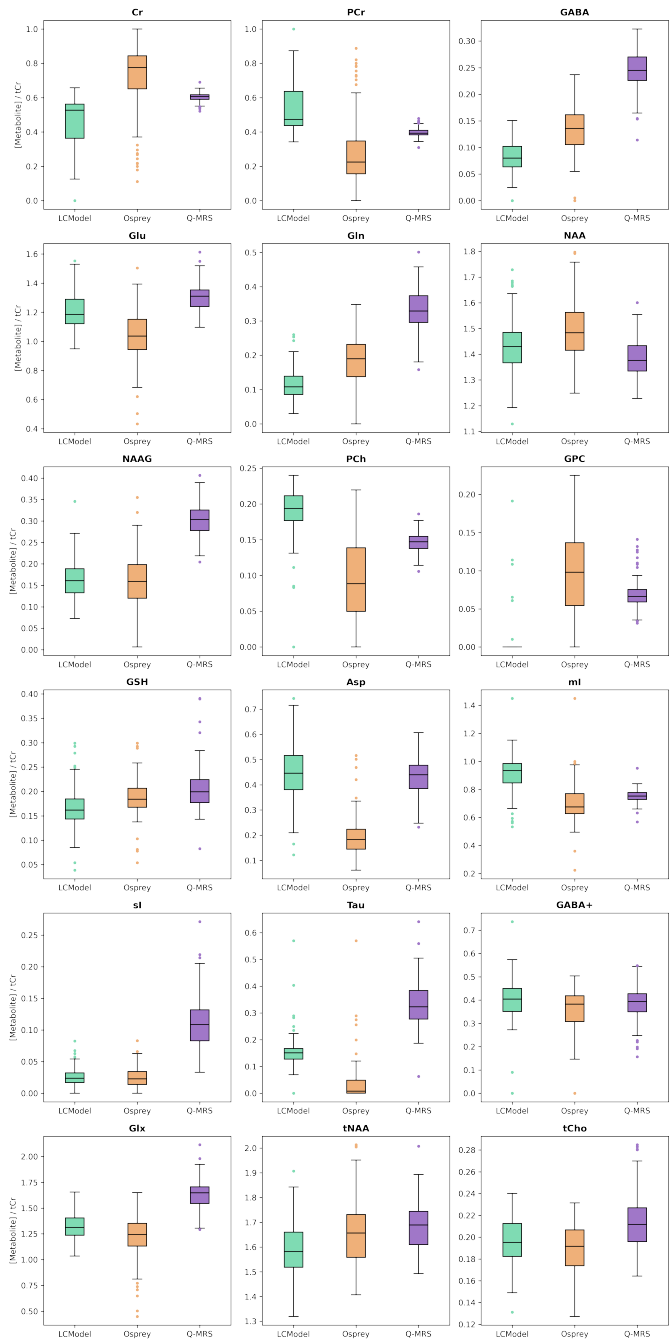

**Fig. S3.** Box plot of estimated metabolite levels.

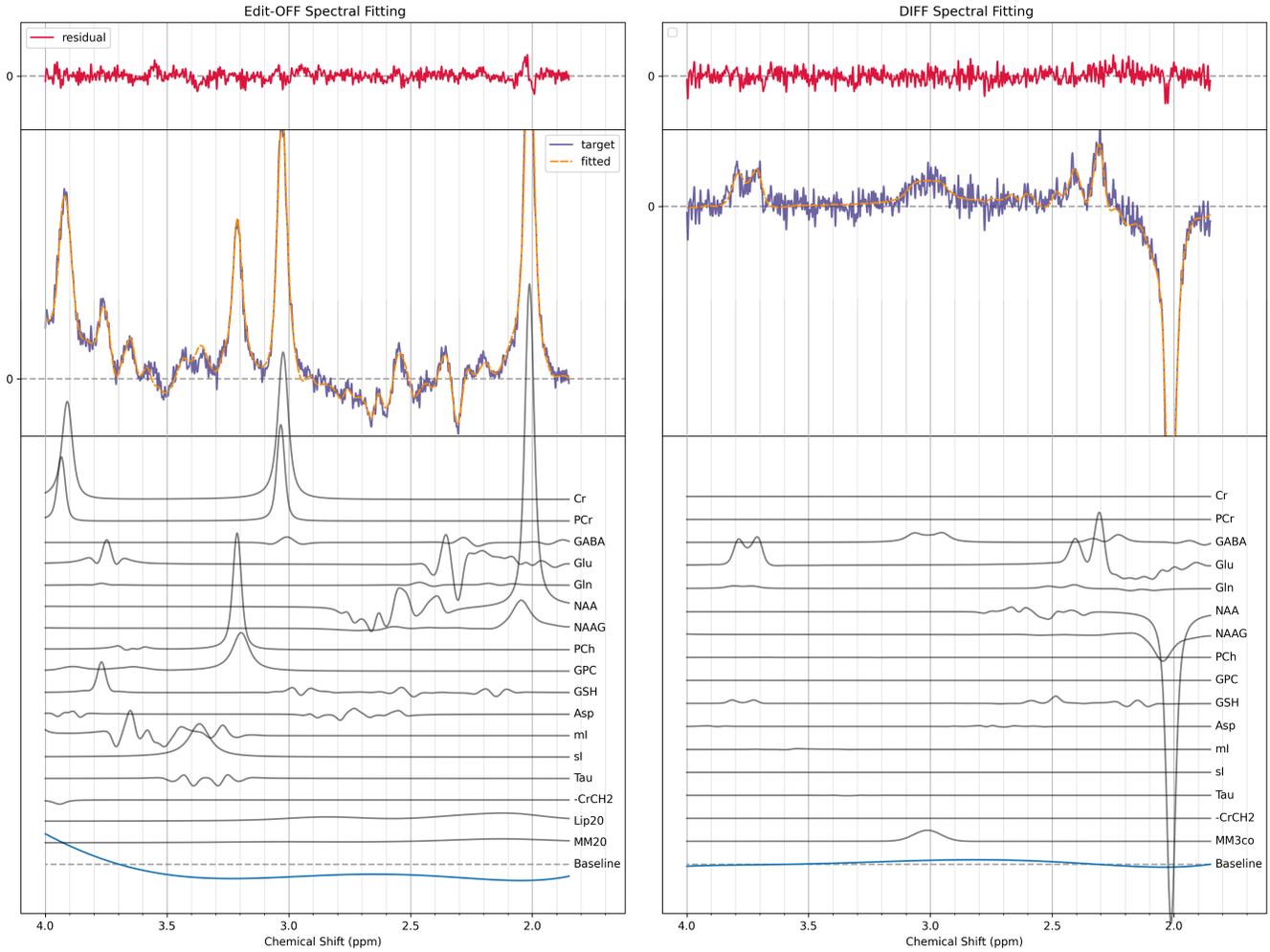

**Fig. S4.** Example of spectral fitting plot for a subject with added noise. The SNRs of the edit-OFF and DIFF spectra are 71.2 and 52.2, respectively.